\documentclass[aps,prd,a4paper,twocolumn,showpacs,nofootinbib]{revtex4-1}

\usepackage{amssymb,amsmath,latexsym,mathrsfs}
\usepackage{graphicx,subfigure}
\usepackage{epsfig}
\usepackage{varioref,xr-hyper}
\usepackage{color}

\begin{document}

\title{Amplitudes of Thermal and Kinetic Sunyaev-Zel'dovich Signals from Small-Scale CMB Anisotropies}

\author{Maria Archidiacono$^1$, Francesco De Bernardis$^2$, Asantha Cooray$^2$, Alessandro Melchiorri$^1$, Alexandre Amblard$^3$, Luca Pagano$^1$, Paolo Serra$^3$}
\affiliation{$^1$Physics Department and INFN, Universita’ di Roma “La Sapienza”, Ple. Aldo Moro 2, 00185, Rome, Italy
$^2$Department of Physics \& Astronomy, University of California, Irvine, CA 92697
$^3$Astrophysics Branch, NASA/Ames Research Center, MS 245-6, Moffett Field, CA 94035}

\begin{abstract}
While the arcminute-scale Cosmic Microwave Background (CMB) anisotropies are due to secondary effects, point sources dominate the total anisotropy power spectrum. At high frequencies the point sources are primarily in the form of dusty, star-forming galaxies. Both {\it Herschel} and Planck have recently measured the anisotropy power spectrum of
cosmic infrared background (CIB)  generated by dusty, star-forming galaxies from degree to sub-arcminute angular scales,
including the non-linear clustering of these galaxies at multipoles of 3000 to 6000 relevant to CMB secondary anisotropy studies.
We scale the CIB angular power spectra to CMB frequencies and interpret the combined WMAP-7 year and arcminute-scale
Atacama Cosmology Telescope (ACT) and South Pole Telescope (SPT) CMB power spectra measurements to constrain the Sunyaev-Zel'dovich (SZ) effects.
Allowing the CIB clustering amplitude to vary, we constrain the amplitudes of thermal and kinetic SZ power spectra at 150 GHz.
\end{abstract}
\maketitle

\section{Introduction}
The Cosmic Microwave Background (CMB) radiation represents a fundamental observable for Cosmology and at present the most powerful one for the investigation of several open questions, such as the nature
of inflation  or primordial non-gaussianity. In two years the Planck satellite \citep{:2011ap} will provide a  measure of the anisotropies of the CMB with a precision never reached before, that will allow a highly precise determination of the standard cosmological parameters as well as major constraints on some non-standard Physics processes.

The observation of CMB anisotropies is affected by several systematics and secondary effects due to the fact that the CMB is not the only source of emission in the microwave frequencies and to the formation of structures
between the observer and the last scattering surface \citep{Tegmark:1995pn,Toffolatti:1997dk,DeZotti:2004mn,Aghanim:2007bt}. The great accuracy of future data requires a compelling description of these effects, in order
to separate the different contributions to the anisotropies and to distinguish primordial and secondary effects (see \cite{Millea:2011pa}). The observable used to extract most of the cosmological information from the CMB is the angular power spectrum $C_\ell$. Secondary effects or unresolved foregrounds provide a contribution to the observed $C_\ell$. In order to obtain an unbiased determination of the cosmological parameters from CMB maps it is necessary to correctly describe possible contaminations. On the other hand, both contaminants and secondary effects themselves contain certain cosmological and astrophysical information, especially on the formation of structure at late-times and the large-scale structure of the Universe,
so that the separation of these components from primordial CMB fluctuations becomes an important science goal on its own.
The Galactic emission and radio point sources are typical examples of foreground contamination in CMB maps.  While the bright sources detected in maps
can be removed with a suitable mask before the estimation of the angular power spectrum, unresolved point sources will contribute to the total anisotropy power spectrum $C_\ell$.

The Sunyaev-Zel'dovich (SZ) effect \citep{sz70}, caused by the Compton scattering of the CMB photons by the electrons in the Universe, is a well-known secondary anisotropy studied by
a variety of experiments. The SZ effect contains cosmological information, since the angular power spectrum of secondary temperature anisotropy arising
from SZ scattering depends  on both the gas distribution in galaxy clusters and on the amplitude of the matter density fluctuations $\sigma_8$ \citep{Barandela_mucket,Komatsu:1999ev,Cooray00,Bond:2002tp}. Since the SZ thermal effect has a unique spectral signature relative to the CMB thermal spectrum, the SZ signal can be distinguished from
primary CMB anisotropies and other foregrounds using observations at multiple frequencies across the SZ null at $\sim 217$ GHz \citep{Cooray:2000xh}.
Such a separation, however, is not feasible with kinetic SZ effect associated with peculiar motions of the electrons scattering the CMB \citep{sz70}
as the signal has the same spectrum as that of the CMB. Even for the SZ thermal effect, in realistic experiments, the
 main obstacles that limit a clear detection of the thermal signal comes from uncertainties in the modeling of the kinetic contribution  and the
difficulty of separating SZ effects from clustered point sources.

The contribution of unresolved point sources and SZ effect is best seen on small angular scales where they dominate the total CMB angular power spectrum.
The use of data at these small angular scales is hence becoming decisive in the analysis of CMB data. In this work we analyze large $\ell$ data (up to $\ell\sim9000$) from the South Pole Telescope (SPT) at $150$ and $220$ GHz \citep{Shirokoff:2010cs} and from Atacama Cosmology Telescope (ACT) \citep{Dunkley:2010ge,Das:2010ga} at $150$ GHz combined with Wilkinson Microwave Anisotropy Probe (WMAP) data after $7$ years of observation \citep{Komatsu:2010fb} to put constraints on the two $SZ$ effects.

In previous studies significant limitations came from uncertainties associated with
clustering of dusty star-forming galaxies (DSFG) that contribute to high-frequency CMB data. Such clustering has now been measured with both {\it Herschel} \citep{Amblard:2011gc}
and Planck \citep{:2011ap} experiments. In the context of CMB studies, {\it Herschel} measurements are most useful as they probe the DSFG clustering down to
sub-arcminute angular scales at scales well matched to arcminute scale CMB experiments while Planck measurements are limited to scales greater than 5$'$ or $\ell < 2000$.
In this work we  describe the clustering of unresolved point sources we used the same template of \cite{Amblard:2011gc}, where the authors reported a detection of both the linear clustering and the non-linear
clustering at a few arcminute scales, corresponding to $\ell \sim 4000$.

We perform a Monte Carlo Markov Chain (MCMC) analysis constraining both the thermal and the kinetic terms of the SZ effect together with the the Poisson and clustering corrections due to unresolved point
sources, including radio sources at lower frequencies such as 150 GHz. In the next Section we recall more details on the contribution of the SZ effect and foregrounds to the CMB anisotropy  power spectra.
Section~3 describes the parametrization and the templates we used to model the foreground and the SZ contamination to the CMB angular anisotropy power spectrum.  In Section~4
we show the results and conclude with a summary.

\section{Parametrizing SZ effect and foregrounds}

In this Section we briefly describe the adopted parametrizations and templates for the Sunyaev-Zel'dovich effect,
unresolved extragalactic point source foregrounds and lensing.

\vspace{0.3cm}
{\it Sunyaev-Zel'dovich thermal and kinetic effect.}

The SZ effect has two different contributions, one from the thermal motion of the electrons (thermal SZ effect - \emph{tSZ}) and one from the bulk motion of the electrons relative to the CMB (kinetic SZ - \emph{kSZ}). The former contribution has a distinct frequency dependence, while the kSZ effect causes only a Doppler shift of the incident CMB spectrum retaining the black body shape. The total SZ signal in a generic direction $\hat{n}$ is then given by (see for example Section~$2$ in \cite{Dunkley:2010ge}):
\begin{equation}\label{SZsignal}
    \Delta T^{\rm SZ}(\nu)=\frac{f(x)}{f(x_0)}\Delta T_0^{\rm tSZ}(\hat{n})+\Delta T^{\rm kSZ}(\hat{n}) \ ,
\end{equation}
with $x=h\nu/k_BT_{CMB}$ and $f(x)=2-x/2\tanh(x/2)$. Here $\Delta T_0^{tSZ}$ is the expected thermal contribution at frequency $\nu_0$. From $f(x)$ it
can easily seen that the thermal SZ effect vanishes at $\sim 218$ GHz.
We model the SZ contributions to the anisotropy angular power spectrum, relative to a template power spectrum, as
\begin {equation}\label{SZ}
D_\ell^{{\rm SZ}, ij}=A_{\rm tSZ}\frac{f(\nu_{i})}{f(\nu_{0})}\frac{f(\nu_{j})}{f(\nu_{0})}D_{0,\ell}^{\rm tSZ}+ A_{\rm kSZ}D_{0,\ell}^{\rm kSZ} \ ,
\end {equation}
where $D_\ell=\ell(\ell+1)C_{\ell}/2\pi$  and $D_{0,\ell}^{i}$ is the template spectrum for either thermal or kinetic SZ.
In this work we consider the SZ templates from \cite{Trac:2010sp}, computed by tracing through a dark matter simulation and processed to include gas in dark matter halos and in the filamentary intergalactic
medium. The thermal SZ template describes the power from tSZ temperature fluctuations from all clusters for a Universe normalized with amplitude of matter fluctuations $\sigma_8 = 0.8$. In particular we use the 'standard' model of \cite{Trac:2010sp}, that was first described in \cite{Sehgal:2009xv}, and assuming a $\sigma_8$ scaling given by $D_{0,\ell}^{tSZ}\propto(\sigma_8/0.8)^{8.1}$ as found in \cite{Trac:2010sp}.  For these templates the reference values at $\ell=3000$ are $D_{0,\ell=3000}^{tSZ}\simeq8.9 \mu K^2$ and $D_{0,\ell=3000}^{kSZ}\simeq2.1 \mu K^2$ 

\vspace{0.3cm}
{\it Foregrounds from unresolved extragalactic point sources.}

The foregrounds contribution to the CMB power spectrum at arcminute angular scales arises essentially from unresolved extragalactic point sources.  These sources provide two contributions, a
Poisson term due to the random discrete distribution and a clustering term
accounting for the large-scale distribution of the sources. We assume the Poisson term as constant in
$C_{\ell}$, modeling it as $D_{\ell}^{\rm Poiss}=A_{\rm Poiss}D_{0,\ell}^{\rm Poiss}$ where $D_{0,\ell}^{\rm Poiss}=(\ell/3000)^2$.
The clustered term can be similarly expressed as $D_{\ell}^{{\rm clust},ij}=A_{\rm clust}(\nu_i,\nu_j)D_{0,\ell}^{\rm clust}$, where $D_{0,\ell}^{\rm clust}$ is the point sources clustering template and $A_{\rm clust}(\nu_i,\nu_j)$ encodes the frequency scaling (see section 3 for further details).
Contribution to point sources comes from radio point sources and dusty star-forming galaxies (DSFG).
At 220 GHz the main point source contribution is mainly DSFGs while at 150 GHz the point sources are primarily radio sources with a synchrotron spectrum. We therefore neglect the clustering of radio sources and assume that the contribution from radio sources is essentially described only by a Poisson behavior.
For clustered DSFGs we adopt the template from \cite{Amblard:2011gc} where the authors reported a detection of both the linear clustering and the excess of clustering associated
with the 1-halo term at arcminute scales. Those  data are from the {\it Herschel} Multi-tiered
Extra-galactic survey (HerMES) \citep{Oliveretal}, taken with the Spectral and Photometric
Imaging Receiver (SPIRE) onboard the {\it Herschel} Space Observatory \citep{Pilbratt:2010mv}.

\vspace{0.3cm}
{\it CMB Lensing}

It is well known that gravitational lensing of CMB anisotropies by large-scale structure tends to increase
the power at small angular scales (see \cite{Hanson:2009kr} for a recent review). A proper calculation of this effect is hence necessary in order to prevent an incorrect estimate of the foregrounds and
SZ parameters. The calculation of lensed CMB spectra out to $\ell=9000$ is prohibitively expensive in computational time. Instead, we approximate the impact of lensing by adding a fixed lensing template
$D_{\ell}^{lens}$ computed by running \texttt{camb} \citep{Lewis:1999bs} with and without the lensing option and taking the difference between these spectra. In this run the cosmological parameters of the $\Lambda$CDM model are fixed at the best fit values WMAP7. In \cite{Shirokoff:2010cs} it has been estimated that the error due to this approximation is less than $0.5\mu$K$^2$ at $\ell>3000$ and is hence negligible with respect to secondary and foregrounds contributions.
The lensing contribution is clearly frequency independent.

\section{Analysis Method and data}

We place constraints on the cosmological parameters and on the SZ and foregrounds parameters using the $7$-years WMAP data in combination with the SPT data at $150$ GHz and at $220$ GHz, and the ACT data at $148$ GHz. The SPT and ACT datasets are necessary to analyze the smaller scales of the power spectrum where point sources and SZ are dominant. For the SPT data we select the single frequency $15\times 15$ blocks from the full $45\times 45$ covariance matrix provided by the SPT collaboration (see \cite{Shirokoff:2010cs}), neglecting the correlation between different frequencies.

We use a $6$-parameter flat-$\Lambda$CDM cosmological model to describe primary CMB anisotropies and reionization:
the baryon and dark matter physical energy densities $\Omega_{b}h^{2}$, $\Omega_{c}h^{2}$, the reionization optical depth $\tau$, the ratio of the sound horizon to the angular diameter distance at the decoupling $\theta$, the
amplitude of the curvature perturbation $A_{s}$ (with flat prior on $\log A_{s}$) and the spectral index $n_{s}$; these two last parameters are both defined at the pivot scale $k_{0}=0.002$ $\rm hMpc^{-1}$. In addition to the standard cosmological parameters we include the SZ and foreground parameters described in the previous section.
We perform a Monte Carlo Markov Chain analysis based on the publicly available package \texttt{cosmomc} \citep{Lewis:2002ah} suitably modified to account for the additional parameters, with a convergence diagnostic based on the Gelman and Rubin statistics. When estimating parameters with point sources and SZ included, the total CMB anisotropy spectra are three, one for each frequency plus the cross-correlation term, because of the frequency dependence of the secondary anisotropies.

In order to study the stability of our results on the assumed parametrization, we perform three different analysis, both with $6$ additional parameters describing SZ effect and foregrounds, but considering different parametrizations.

\vspace{0.3cm}
{\it First case: "run1".}

In the first case, that we define as "run1" in what follows, we consider the SZ effect parameters $A_{\rm tSZ}$ and $A_{\rm kSZ}$, the Poissonian contribution $A_{\rm Poiss}^{150}$ and $A_{\rm Poiss}^{220}$ and the Poisson contribution for the 150$\times$220 GHz cross-correlation $A_{\rm Poiss}^{X}$.
The use of the $A_{\rm Poiss}^{X}$ extra parameter for the cross correlation of Poisson point source can be justified from
the possibility that the contribution at one single frequency comes from more than one point source population and that the
two channels are not fully correlated. This possibility has not been considered in previous analyses and is therefore important to evaluate the impact of this assumption.
Finally we consider a {\it single} clustered point sources parameter, $A_{\rm clust}$,  scaling the contribution at  different frequencies using the relation of \cite{Gispert:2000np}:

\begin{eqnarray*}
I_{\nu}=8.80\times10^{-5}\left(\nu/\nu_0\right)P_\nu\left(13.6K\right) \ ,
\end{eqnarray*}
with $\nu_0 = 100 {\rm cm} ^{-1}$, following recent results from Planck \citep{:2011ap}. In what follows we
refer to this scaling as "Gispert" scaling.

In summary, the spectra in "run1" are defined as:

\begin{eqnarray*}
D_{\ell}(150)&=&D_{\ell}^{lens}+A_{\rm tSZ}D_{0,\ell}^{\rm tSZ}+A_{\rm kSZ}D_{0,\ell}^{\rm kSZ} \nonumber \\
&& +A_{\rm clust}D_{0,\ell}^{\rm clust150} +A_{\rm Poiss}^{150}D_{0,\ell}^{\rm Poiss} \\
D_{\ell}(220)& = & D_{\ell}^{lens}+A_{\rm kSZ}D_{0,\ell}^{\rm kSZ}\nonumber \\
&& +A_{\rm clust}D_{0,\ell}^{\rm clust220}+A_{\rm Poiss}^{220}D_{0,\ell}^{\rm Poiss} \\
D_{\ell}({\rm 150\times220})&=&D_{\ell}^{lens}+A_{\rm kSZ}D_{0,\ell}^{\rm kSZ}\nonumber \\
&& +A_{\rm clust}D_{0,\ell}^{\rm clustcross}+A_{\rm Poiss}^{\rm cross}D_{0,\ell}^{\rm Poiss}
\end{eqnarray*}

The thermal SZ effect is negligible at $220$ GHz. The contribution of the thermal SZ effect to the cross-correlated power spectrum may not vanish in presence of a spatial correlation between IR sources and the clusters that cause the thermal SZ. Nevertheless as showed in \cite{Shirokoff:2010cs} the effect of this correlation is negligible for the SPT data (see par. 7.4 in \cite{Shirokoff:2010cs} for further details). We hence do not consider this contribution when fitting the data.

\vspace{0.3cm}
{\it Second case: "run2".}

In the second analysis, to which in what follows we refer as "run2", we assume full correlation between the Poisson point sources signal at $150$ and $220$ GHz as done in previous analyses, i.e. we fix the cross amplitude of Poisson point sources at the square root of the product of the amplitudes at $150$ and $220$, $A_{\rm Poiss}^{X} =\sqrt{A_{\rm Poiss}^{150}A_{\rm Poiss}^{220}}$.
Moreover, we don't scale the clustered point sources template and we use instead  two different parameters for $150$ and $220$ GHz.
This second analysis is more similar to the one presented in \cite{Shirokoff:2010cs}, however we point out that while
here we consider the amplitudes at different frequencies as free parameters, \cite{Shirokoff:2010cs} considered
the amplitudes at one single frequency and one common frequency spectral index for clustered and point sources
as free parameters.

In this second case the foreground spectra are defined as:

\begin{eqnarray*}
D_{\ell}(150)&=&D_{\ell,150}^{lens}+A_{\rm tSZ}D_{0,\ell}^{\rm tSZ}+A_{\rm kSZ}D_{0,\ell}^{\rm kSZ}\nonumber\\
&& +A_{\rm clust150}D_{0,\ell}^{\rm clust}+A_{\rm Poiss}^{150}D_{0,\ell}^{\rm Poiss} \\
D_{\ell}(220)& = & D_{\ell,220}^{lens}+A_{\rm kSZ}D_{0,\ell}^{\rm kSZ}\nonumber \\
&& +A_{\rm clust220}D_{0,\ell}^{\rm clust}+A_{\rm Poiss}^{220}D_{0,\ell}^{\rm Poiss} \\
D_{\ell}({\rm 150\times220})&=&D_{\ell,X}^{lens}+A_{\rm kSZ}D_{0,\ell}^{\rm kSZ} \nonumber \\
&& +\sqrt{A_{\rm clust150}A_{\rm clust220}} D_{0,\ell}^{\rm clust} \nonumber \\
&& +\sqrt{A_{\rm Poiss150}A_{\rm Poiss220}}D_{0,\ell}^{\rm cross}
\end{eqnarray*}

\vspace{0.3cm}
{\it Third case: "run3".}

Finally we combine $150$ GHz data of SPT and ACT, using separate parameters for ACT and SPT both for clustered and Poisson point sources, to account for the different masking thresholds of the point sources. In this case we have:

\begin{eqnarray*}
D_{\ell}(150)&=&D_{\ell,150}^{lens}+A_{\rm tSZ}D_{0,\ell}^{\rm tSZ}+A_{\rm kSZ}D_{0,\ell}^{\rm kSZ}\nonumber \\
&& +A_{\rm clustACT}D_{0,\ell}^{\rm clust}+A_{\rm PoissACT}^{150}D_{0,\ell}^{\rm Poiss}\nonumber \\
&& +A_{\rm clustSPT}D_{0,\ell}^{\rm clust}+A_{\rm PoissSPT}^{150}D_{0,\ell}^{\rm Poiss} \nonumber
\end{eqnarray*}

\begin{table*}[!]
\begin{center}
\begin{tabular}{lrrrr}
\hline
\hline
                       &WMAP7+                         & WMAP7+               &WMAP7+                           &WMAP7+\\

                       &    SPT (run1, kSZ free )      &   SPT (run1)         &SPT(run2, kSZ free)              & SPT (run2)\\
\hline
 $ 10^2\Omega_b h^2$   & $2.267\pm0.049$               & $2.268\pm0.051$      &    $2.264\pm0.049$                      & $2.269\pm0.049$  \\
 $ \Omega_c h^2$       & $0.113\pm0.005$               & $0.113\pm0.004$      & $0.1126\pm0.0052$                       & $0.1127\pm0.0052$ \\
 $ \tau$               & $0.090\pm0.015$               & $0.089\pm0.015$      & $0.089\pm0.015$                         & $0.090\pm0.014$  \\
 $ n_s$                & $0.973\pm0.013$               & $0.973\pm0.013$      & $0.972\pm0.013$                         & $0.972\pm0.012$  \\
 $ln(10^{10} A_s)$     & $3.18\pm0.04$                 & $3.18\pm0.04$        & $3.18\pm0.045$                          & $3.18\pm0.04$    \\
 $ \Omega_m$           & $0.278\pm0.028$               & $0.279\pm0.029$      & $0.276\pm0.029$                         & $0.276\pm0.028$  \\
 $\sigma_8$            & $0.823\pm0.028$               & $0.825\pm0.028$      & $0.820\pm0.0272$                        & $0.821\pm0.026$  \\
 $A_{\rm tSZ}$         & $0.24\pm0.17$                 & $0.25\pm0.16$        & $0.33\pm0.23$                           & $0.52\pm0.22$    \\
 $A_{\rm kSZ}$         & $1.3\pm0.9$                   & $\left[1\right]$     & $2.7\pm1.4$                             & $\left[1\right]$ \\
 $A_{\rm clust}$       & $1.05\pm0.19$                 & $1.08\pm0.14$        & $ - $                                   & $ - $            \\
 $A_{\rm clust150}$    & $ - $                         & $ - $                & $0.44\pm0.27$                           & $0.66\pm0.26$   \\
 $A_{\rm clust220}$    & $ - $                         & $ - $                & $8.2\pm1.7$                             & $8.7\pm1.5$ \\
\hline
 $D_{\ell 3000}^{\rm tSZ}$($\mu$K$^2$)       & $2.2\pm1.5$          & $2.3\pm1.4$        & $2.9\pm2.0$           & $4.7\pm2.0$    \\
 $D_{\ell 3000}^{\rm kSZ}$($\mu$K$^2$)       & $2.7\pm1.9$          & $\left[2.05\right]$   & $5.5\pm3.0$           & $\left[2.05\right]$ \\
 $D_{\ell 3000}^{\rm clust150}$($\mu$K$^2$)  & $6.05\pm1.06$        & $6.26\pm0.82$      & $2.51\pm1.60$         & $3.81\pm1.53$   \\
 $D_{\ell 3000}^{\rm clust220}$($\mu$K$^2$)  & $39.11\pm6.79$       & $40.63\pm5.08$     & $47.33\pm9.78$        & $50.47\pm9.17$ \\
 $D_{\ell 3000}^{\rm Poiss150}$($\mu$K$^2$)   & $10.03\pm0.67$      & $10.1\pm0.7$       & $10.38\pm0.63$        &  $10.33\pm0.67$  \\
 $D_{\ell 3000}^{\rm Poiss220}$($\mu$K$^2$)   & $79.5\pm4.8$        &  $80\pm5$          & $77.89\pm4.49$        &  $76.5\pm4.0$   \\
 $D_{\ell 3000}^{\rm Poisscross}$($\mu$K$^2$) & $26.8\pm1.4$        &  $26.8\pm1.4$      & $ - $                 & $ - $          \\
 \hline
\hline
\end{tabular}
\caption{Mean values and $68\%$ error bars from SPT data at 150 and 220 GHz. Run1 case is  with only one DSFG clustering
amplitude allowed to vary and frequency scaling fixed from \cite{Gispert:2000np}, consistent with Planck \citep{:2011ap}.
Run2 case is with two DSFG clustering amplitudes allowed to vary, without frequency scaling.}
\label{spt_tab}\vspace{1cm}

\begin{tabular}{lrr}
\hline
\hline
                       &WMAP7+                                   & WMAP7+\\
                       &    SPT+ACT (kSZ free )                  & SPT+ACT \\
\hline
 $ 10^2\Omega_b h^2$   &    $2.232\pm0.047$                      & $2.234\pm0.046$  \\
 $ \Omega_c h^2$       & $0.1121\pm0.0050$                       & $0.1124\pm0.0053$ \\
 $ \tau$               & $0.086\pm0.014$                         & $0.086\pm0.015$  \\
 $ n_s$                & $0.964\pm0.012$                         & $0.964\pm0.012$  \\
 $ln(10^{10} A_s)$     & $3.20\pm0.043$                          & $3.19\pm0.04$    \\
 $ \Omega_m$           & $0.274\pm0.027$                         & $0.275\pm0.028$  \\
 $\sigma_8$            & $0.812\pm0.0255$                        & $0.813\pm0.027$  \\
 $A_{\rm tSZ}$             & $0.34\pm0.25$                       & $0.38\pm0.24$    \\
 $A_{\rm kSZ}$             & $1.6\pm1.1$                         & $\left[1\right]$ \\
 $A_{\rm clustact}$        & $0.66\pm0.56$                       & $0.75\pm0.59$   \\
 $A_{\rm clustspt}$        & $0.66\pm0.43$                       &  $0.77\pm0.41$  \\
\hline
 $D_{\ell 3000}^{\rm tSZ}$($\mu$K$^2$)             & $3.1\pm2.3$                  & $3.5\pm2.2$    \\
 $D_{\ell 3000}^{\rm kSZ}$($\mu$K$^2$)             & $3.2\pm2.3$                  & $\left[2\right]$ \\
 $D_{\ell 3000}^{\rm clustact}$($\mu$K$^2$)       & $3.9\pm3.2$                  & $4.2\pm3.2$   \\
 $D_{\ell 3000}^{\rm Poissact}$($\mu$K$^2$)    & $13.4\pm2.4$         & $13.5\pm2.5$ \\
 $D_{\ell 3000}^{\rm clustspt}$    ($\mu$K$^2$)            & $3.8\pm2.5$          &  $4.5\pm2.4$  \\
 $D_{\ell 3000}^{\rm Poissspt}$($\mu$K$^2$)    & $10.2\pm0.8$         &  $10.2\pm0.8$   \\
 \hline
\hline
\end{tabular}
\caption{Mean values and $68\%$ error bars from ACT data combined with SPT data at $150$ GHz}
\label{spt_act_tab}\vspace{1cm}
\end{center}
\end{table*}

\section{Results}

In Table~\ref{spt_tab} we report the mean values of the cosmological parameters and their $68\%$ C.L. uncertainty from SPT data at $150$ and $220$ GHz for the "run1" and "run2" analyses, while
in Table~\ref{spt_act_tab} we list the mean values of the cosmological parameters and their $68\%$ C.L. uncertainty from ACT data ("run3") combined with SPT data at $150$ GHz.
In order to facilitate the comparison with other works present in the literature we also translate the constraints on
the foregrounds amplitudes in to the foreground power spectrum at $\ell=3000$, $D_{\ell=3000}$.
Since a significant correlation exists between thermal and kinetic SZ and since the kinetic SZ is predicted to be
small, we also perform an analysis by fixing $D_{\ell=3000}^{\rm kSZ}=2 \, \mu K^2$.

We find that for the ``run1'' case the thermal SZ anisotropy amplitude is $D_{\ell=3000}^{\rm tSZ}=2.2\pm1.5 \, \mu K^2$ . While a  $\sim  1 \sigma$ indication for SZ is present our result is less significant than the one reported by \cite{Shirokoff:2010cs} with $D_{\ell=3000}^{\rm tSZ}=3.2\pm1.3 \, \mu K^2$ i.e. with a thermal SZ detection at at more than two standard deviations. The result on the kinetic SZ component are compatible, with
$D_{\ell=3000}^{\rm kSZ}=2.7\pm1.9 \, \mu K^2$ at $68 \%$ c.l. from our analysis to be compared with
$D_{\ell=3000}^{\rm kSZ}=2.4\pm2.0 \, \mu K^2$ from \cite{Shirokoff:2010cs}.

\begin{figure} [h!]
\includegraphics[width=\columnwidth]{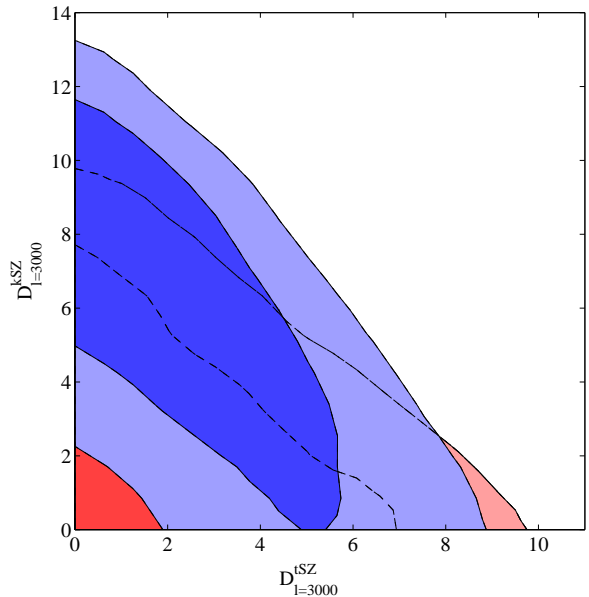}
  \caption{Joint two-dimensional posterior probability contours showing $68\%$ and $95\%$ C.L. constraints on $D_{\ell=3000}^{\rm ksz}$ and $D_{\ell=3000}^{\rm tSZ}$ from ACT $150$ GHz data (red) and SPT all frequencies data (blue) for the run2 case.}
\label{sz}
\end{figure}

Although the point sources and SZ parameters do not show significant degeneracies with cosmological parameters (see also \cite{Serra08}), a strong correlation exists between $A_{\rm tSZ}$ and $A_{\rm kSZ}$ and to a smaller extent between $A_{\rm tSZ,kSZ}$ and $A_{\rm clust}$. This can be seen in Figure~\ref{sz} where we show the $2-D$ likelihood constraints in the plane $D_{\ell=3000}^{\rm kSZ}-D_{l=3000}^{\rm tSZ}$ for the "run2" and "run3" case. Fixing the kSZ term sligthly improves the detection for the
thermal SZ with $D_{\ell=3000}^{\rm tSZ}=2.3\pm1.4 \, \mu K^2$ in ``run1'' but still with less significance than the one in
\cite{Shirokoff:2010cs} where a value of $D_{\ell=3000}^{\rm tSZ}=3.5\pm1.0 \, \mu K^2$ is reported.

Based on the degeneracy direction of  Figure~\ref{sz},	we constrain the sum of the SZ effects at
 $\ell=3000$ to be $D^{\ell=3000}_{\rm tSZ}+0.5 D^{\ell=3000}_{\rm kSZ}= 3.5 \pm 1.8$ $\mu$K$^2$ to be compared $4.5 \pm 1.0$ $\mu$K$^2$ of \cite{Shirokoff:2010cs} .
These amplitudes are consistent but, again, the significance of the detection is worse than \cite{Shirokoff:2010cs} who found this sum to be $4.5 \pm 1.0$ $\mu$K$^2$.

The small discrepancy with the results presented in \cite{Shirokoff:2010cs} comes essentially
from the different parametrization used.
Adopting a more similar parametrization as in the case of "run2" we found $D_{\ell=3000}^{\rm tSZ}=2.9\pm2.0 \, \mu K^2$, $D_{\ell=3000}^{\rm kSZ}=5.5\pm3.0 \, \mu K^2$ at $68 \%$ c.l., $D^{\ell=3000}_{\rm tSZ}+0.5 D^{\ell=3000}_{\rm kSZ}= 5.6 \pm 2.6$ $\mu$K$^2$, yielding a detection for the thermal SZ with higher significance.
In case of fixed kSZ we obtain $D_{\ell=3000}^{\rm tSZ}=4.7\pm2.0 \, \mu K^2$, again a more significant detection in better agreement with \cite{Shirokoff:2010cs}.

The different assumptions in the frequency scaling of the clustered point sources component in "run1" and "run2"
is the main explanation for the difference in the results.
 In ``run1'',  taking into account Gispert scaling \citep{Gispert:2000np}, we find $D_{\ell=3000}^{\rm clust220}=39.11\pm6.79 \, \mu K^2$, while in ``run2'', when the amplitude of the clustering point sources is allowed to vary, we have   $D_{\ell=3000}^{\rm clust220}=47.33\pm9.78 \, \mu K^2$, that is more consistent with the corresponding value of $D_{\ell=3000}^{\rm clust220}=57\pm9$ reported in \cite{Shirokoff:2010cs}. A small tension therefore exists between the Gispert scaling and the data at $220$ GHz, resulting also in a worse determination of the thermal SZ signal.
The use of a different parametrization of the point source (just amplitudes in our
case while \cite{Shirokoff:2010cs} varies one amplitude and one spectral index per component) can explain
the remaining differences.

Concerning the Poisson point sources component at $150$ GHz, our results  are different when 
directly compared to those from \cite{Shirokoff:2010cs}, both in ``run1'' and ``run2''. At $150$ GHz we find $D_{\ell=3000}^{\rm Poiss150}=10.03\pm0.67 \, \mu K^2$ in ``run1'' and $D_{\ell=3000}^{\rm Poiss150}=10.38\pm0.63 \, \mu K^2$ in ``run2'', while the value in \cite{Shirokoff:2010cs} is $D_{\ell=3000}^{\rm Poiss150}=7.4\pm0.6 \, \mu K^2$. 
This difference is explained in straightforward terms if we take in account that in \cite{Shirokoff:2010cs} radio galaxies are included in their ``baseline model'' with an amplitude $D_{\ell=3000}^{\rm r}=1.28 \, \mu K^2$ with a $15\%$ uncertainty. Clustering of radio galaxies is negligible, so this radio galaxies term is a Poisson like term of the form $ \propto \ell^2$. Adding this component, our Poisson amplitudes are consistent with those reported by \cite{Shirokoff:2010cs} within $1\sigma$. In ``run1'' at $150$ GHz we find $D_{\ell=3000}^{\rm Poiss150}=10.03\pm0.67 \, \mu K^2$, while in \cite{Shirokoff:2010cs} the correspondent total Poisson contribution at $\ell=3000$ is about $(8.68 \pm 0.69) \, \mu K^2$. At $220$ GHz we find   $D_{l=3000}^{\rm Poiss220}=79.5\pm4.8 \, \mu K^2$ in ``run1'' and $D_{\ell=3000}^{\rm Poiss220}=77.89\pm4.49 \, \mu K^2$ in ``run2'', while the value in \cite{Shirokoff:2010cs} is $D_{\ell=3000}^{\rm Poiss220}=71\pm5 \, \mu K^2$.

We can therefore conclude that the current results presented in the literature on the
amplitude of the secondary anisotropies should be considered with great care since there is
a clear dependence on the parametrization used, on the frequency scaling adopted and on the
assumed templates. We stress that, a part for small discrepancies imputable to differences in the parameterization, all our results for the SZ amplitudes from the analysis of SPT data both for our "run1" and "run2" cases, are substantially consistent with the analysis of the same data made by \cite{Shirokoff:2010cs},  even if we are finding less tight constraints. Our results hence compare in the same way to the recent predictions of tSZ power made by the models of \cite{Shaw2010}, \cite{Trac:2010sp} and \cite{Battaglia2010} confirming that these models overestimate the power of the tSZ signal, as already found in \cite{Shirokoff:2010cs}.


\begin{figure}[h!]
 \includegraphics[width=\columnwidth]{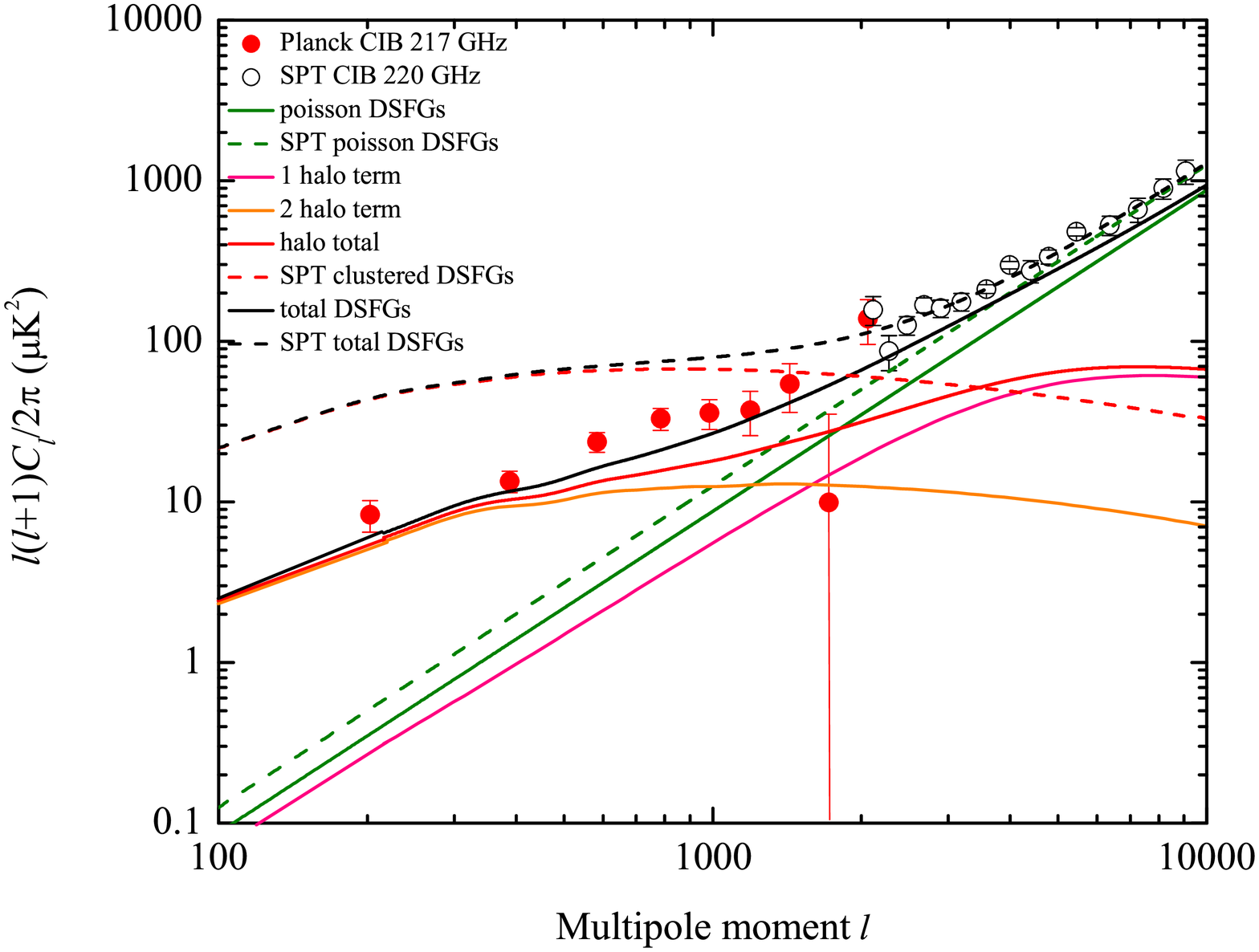}
  \caption{Planck DSFG clustering data (red points) at $217$ GHz and SPT data at $220$ GHz (white points) compared with the combination (solid black line) of the Poisson term (green line) and clustering term (red) of unresolved point sources by scaling the best-fit model to measurements made with {\it Herschel} at 350 $\mu$m to 217 GHz using the frequency scaling of
\cite{Gispert:2000np}. We show the 1-halo (pink) and 2-halo (orange) contributions to the clustering term following \cite{Amblard:2011gc}.
The dashed lines are the $220$ GHz SPT DSFG power spectrum components from \cite{Hall:2009rv}, which resulted in an overestimate of Planck DSFG clustering at $\ell < 3000$.
}\label{CIB}
\end{figure}

In Figure \ref{CIB}  we show the recent CIB power spectra data of the Planck collaboration \citep{:2011ap} at 217 GHz and small angular scale CMB power spectrum data from SPT, at 220 GHz, with a comparison to scaled measurements from \cite{Amblard:2011gc}. The {\it Herschel} model is shown in terms of the 1-halo and 2-halo contributions to the total power spectrum. For reference, we also show the model used by \cite{Hall:2009rv} at 220 GHz to describe the clustering of DSFGs, which overestimated the power at tens of arcminute angular scales and above relative to {\it Herschel} and Planck DSFG clustering measurements. \cite{Hall:2009rv} used a linear model to analyze their data. At small angular scales non-linear effects are not negligible and using a linear model to interpret the data may lead to a wrong determination of the bias and hence to an overestimation of the power at larger angular scales (see also discussion in \cite{:2011ap}). Instead our model shows a good fit of both Planck and SPT CIB. 

\begin{figure*}[ht!]
  \begin{center}
\includegraphics[width=2 \columnwidth]{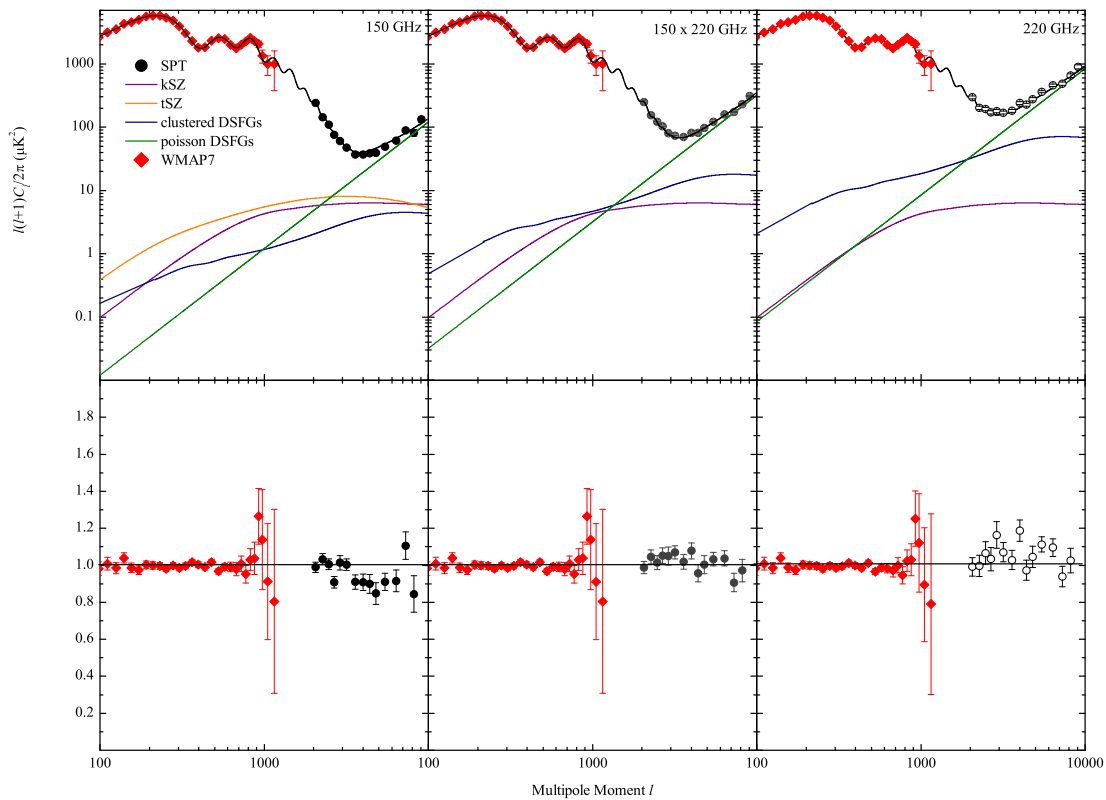}
\end{center}
  \caption{Contribution to the angular anisotropy power spectrum from point sources and from SZ effect for the best fit model of the WMAP7+SPT analysis. Left panel is $150$ GHz, middle $220$ GHz and right panel shows the cross spectra. kSZ term is the orange solid line and the tSZ term at $150$ GHz is the purple line. Green lines are the Poisson terms and blue lines are the clustering contributions. The black lines are the total best fit power spectra. Black dots are SPT data and red squares are WMAP7 data. The bottom panels show the residual relative to the total model, including primordial CMB and best-fit secondary anisotropy amplitudes.}\label{spectra}
 \end{figure*}

In Figure~\ref{spectra} we show the best fit models for each component compared with the SPT and WMAP7 data. In the ``run1'', when only one amplitude of clustered DSFGs is allowed to vary when fitting the all frequencies SPT data combined with WMAP7 data, we find that $A_{\rm clust}=1.05\pm0.19$. This suggests that the combination of \cite{Amblard:2011gc} model and the frequency scaling for the mean CIB is a good fit of the DSFG clustering at lower CMB frequencies. Higher precision CMB power spectra at $150$, $220$ and $350$ GHz and a direct cross-correlation of {\it Herschel}-SPIRE maps against the CMB will be necessary to study if fluctuations scale with frequency as the mean CIB intensity and to improve overall constraints on secondary anisotropies.

\section{Conclusions}

In conclusion, we provided a new analysis of the foreground contribution to the CMB data making use of the
latest ACT and SPT results. Our work is complementary to those presented by the SPT and ACT experimental teams
since we use a different parametrization of the foregrounds contribution and different templates.
The foreground contribution from Poisson point sources at $220$ and $150$ GHz is detected with very high
significance (at more than $\sim 15$ standard deviations) with no particular dependence on the
parametrization used ("run1" and "run2" cases are giving very consistent results).
The contribution from clustered point sources is also well detected at $220$ GHz. We have found that current CMB
data favours a larger contribution at this frequency than the one expected by the Gispert frequency scaling once
the data is normalized at $150$ GHz.
The thermal SZ component is detected at a level slightly above the two standard deviations. However a different
parametrization of the components and the assumption of the Gispert scaling could bring this detection to about
one standard deviation. The correlation with the kinetic SZ term is present in the data despite the
multi-frequency approach. More data at more frequencies are clearly needed to establish a strong detection
of the SZ term.

While our constraints do not improve the results in the literature, we have made a significant addition to prior studies by firmly establishing the power spectrum of DSFGs that dominate the arcminute scale CMB anisotropies at 220 GHz and higher frequencies. This comes from the recent {\it Herschel} results combined with Planck-confirmed frequency spectrum for the CIB mean intensity. In future, additional improvements will come from directly cross-correlating the CMB maps against high-resolution CIB maps from {\it Herschel}; for this a {\it Herschel}-SPIRE survey at the same large areas as CMB surveys will become useful \citep{Cooray10}.

\section{Acknowledgments}
It is a pleasure to thank Erminia Calabrese for useful suggestions and comments. MA thanks the group at UCI for
hospitality while this research was conducted. We thank NSF AST-0645427 and NASA NNX10AD42G for support.

  \end{document}